\documentclass[pre,aps,onecolumn,superscriptaddress,notitlepage,nofootinbib]{revtex4-1}
\usepackage{amssymb,epsfig,amsmath,bm,subfigure,graphicx,textcomp,url,float,color,cases}
\usepackage{epsfig}
\usepackage{textcomp}
\usepackage[normalem]{ulem}

\usepackage[colorlinks=true, urlcolor=blue, anchorcolor=blue, citecolor=blue,filecolor=blue,linkcolor=blue,menucolor=blue
]{hyperref}

\usepackage{color}

\usepackage[titletoc,title]{appendix}

\begin{document}
\title{Anomalous scaling and phase transition in large deviations of dynamical observables 
of stationary Gaussian processes}

\author{Alexander Valov}
\email{aleksandr.valov@ib.edu.ar}
\affiliation{Instituto Balseiro, Univ. Nacional de Cuyo, Av. Bustillo, 9500, Argentina}

\author{Baruch Meerson}
\email{meerson@mail.huji.ac.il}
\affiliation{Racah Institute of Physics, Hebrew University of Jerusalem, Jerusalem 91904, Israel}

\begin{abstract}
We study large deviations, over a long time window $T \to \infty$, of the dynamical observables $A_n = \int_{0}^{T} x^n(t), dt$, $n=3,4,\dots$, where $x(t)$ is a centered stationary Gaussian process in continuous time. We show that, for short-correlated processes the probability density of $A_n$ exhibits an anomalous scaling $P(A_n,T) \sim \exp[-T^{\mu} f_n(\Delta A_n T^{-\nu})]$ at $T\to \infty$ while keeping $\Delta A_n T^{-\nu}$ constant. Here $\Delta A_n$ is the deviation of $A_n$ from its ensemble average. The anomalous exponents $\mu$ and $\nu$ depend on $n$ and are smaller than $1$, whereas the rate function $f_n(z)$ exhibits a first-order dynamical phase transition (DPT) which resembles  condensation transitions observed in many systems. The same type of anomaly and DPT, with the same $\mu$ and $\nu$, was previously uncovered for the Ornstein-Uhlenbeck process -- the only stationary Gaussian process which is also Markovian. We also uncover an anomalous behavior and a similar DPT in the \emph{long-correlated} Gaussian processes. However, the anomalous exponents $\mu$ and $\nu$ are determined in this case not only by $n$ but also by the power-law long-time decay $\sim |t|^{-\alpha}$ of the covariance function. The different anomalous scaling behavior is a consequence of a faster-than-linear scaling with $T$ of the variance of $A_n$. Finally, for sufficiently long-ranged correlations, $\alpha<2/n$,  the DPT disappears, giving way to a smooth crossover between the regions of typical, Gaussian fluctuations and  large deviations. The basic mechanism behind the DPT is the existence of strongly localized optimal paths of the process conditioned on very large  $A_n$ and coexistence between the localized and delocalized paths of the conditioned process. Our theoretical predictions are corroborated by replica-exchange Wang-Landau simulations where we could probe probability densities down to $10^{-200}$.

\end{abstract}

\maketitle
\nopagebreak

\section{Introduction}

Fluctuations of ``dynamical observables", which describe the cumulative effect of the histories of stochastic processes over a long time 
window $T$, is an important subject of nonequilibrium statistical mechanics and probability theory \cite{Touchette2009}. 
Theoretical description of dynamical observables is challenging because they defy standard tools of 
equilibrium statistical mechanics: in  particular, by exhibiting nonequilibrium behaviors  even
when the underlying system is in thermal equilibrium.  Furthermore,  in the limit of $T\to \infty$ the full probability distribution of dynamical observables -- which includes large deviations \cite{Touchette2009,Oono,Dembo,denHollander} -- can exhibit singularities which,  by analogy with equilibrium phase transitions, are called dynamical 
phase transitions (DPTs).

A simple but important class of dynamical observables deals with the quantities
\begin{equation}\label{An}
A_n=\int_{0}^{T} x^n(t) dt\,,\quad n=1,2, \dots\,,
\end{equation}
where $x(t)$ is a centered stationary stochastic process. In standard cases  fluctuations of these quantities obey a simple large-deviations principle where, in the limit of $T\to \infty$,
the probability distribution of $A$ behaves as
\begin{equation}\label{LDP}
P(A_n,T) \sim e^{-T \Phi_n(\Delta A_n/T)},
\end{equation}
where $\Delta A_n$ is the deviation of $A_n$ from its mean \cite{Touchette2018}. The rate function $\Phi_n(y)$ is usually convex, and it vanishes at $y=0$. 
If the process is Markovian, $\Phi_n(y)$ can be obtained from the Donsker–Varadhan (DV) theory, where the calculations reduce to determining the largest eigenvalue of a modified (tilted) generator of the process $x(t)$; see Ref. \cite{Touchette2018}  for an accessible exposition. 
The DV theory, however, breaks down when the spectrum of the tilted generator is not bounded from above \cite{NT2018,denHollander2019,Buisson2020}. As a result,
an anomalous scaling
\begin{equation}\label{anomLDP}
P(A_n,T) \sim e^{-T^{\mu} f_n(\Delta A_n/T^{\nu})}\,,
\end{equation}
arises in the limit of $T\to \infty$, while keeping $\Delta A_n/T^{\nu}$ constant \cite{Smith2022}. The anomalous exponents $\mu$ and $\nu$ differ from the standard values $\mu=\nu=1$, and they depend on the specific system and on $n$. Anomalous scaling of this type is observed, at $n>2$,  already for the classic
Ornstein-Uhlenbeck process, which describes an overdamped motion of a Brownian particle in a quadratic potential \cite{OU}. In this case one obtains \cite{Smith2022}
\begin{equation}\label{munu}
\mu = \frac{2}{2n-2} \quad \text{and} \quad \nu = \frac{n}{2n-2}\,.
\end{equation}
The anomalous scaling behavior is closely related to a change in the character of the most likely history of the process conditioned on a value of $A_n$: from contributions spread uniformly in time to localized, instanton-like events that produce a macroscopic contribution to $A_n$ \cite{NT2018,Meerson2019,NT2022,Smith2022,Bassanoni}.  Remarkably, the rate function $f_n(z)$ in Eq.~(\ref{anomLDP}) exhibits a first-order dynamical phase transition (DPT) \cite{Smith2022} which is similar to ``condensation transitions" observed in many other settings, see e.g. \cite{Brosset2007,Mori2021a,Mori2021b,SmithMajumdar2022,Smith2024}. The mechanism behind the first-order DPT is the coexistence (at different times) of two different types of the time histories of the system -- delocalized and localized in time -- which dominate the contribution to $A_n$ \cite{Smith2022}. This mechanism manifests itself in the expression for rate function  $f_n(z)$ in Eq.~(\ref{anomLDP}) \cite{Smith2022}:
\begin{equation}
\label{f} 
    f_n(y) = \min\limits_{z\in [0,y]} F_n(y,z),
\end{equation}
where
\begin{equation}
\label{F}
    F_n(y,z) = \beta_n (y-z)^2 + c_n z^{2/n},\quad n>2,
\end{equation}
where $\beta_n >0$, $c_n>0$, and $n=2,3,\dots$. The first, quadratic term on the right-hand side of Eq.~(\ref{F}), with
\begin{equation}\label{betan}
\beta_n \simeq \frac{T}{2 \text{Var}(A_n)} ,
\end{equation}
describes the contribution of the (typical) Gaussian fluctuations of $A_n$ around the mean $\langle A_n \rangle$. 
For the OU process, the variance of $A_n$ grows as $T$ at large $T$, therefore $\beta_n$ is time-independent. 

The second term in Eq.~(\ref{F}) describes the contribution of a strongly localized, instanton-like optimal path of the process which dominates the probability density of large $A_n$ \cite{NT2018}. 
The first-order DPT appears, at $T\to \infty$, at the point
\begin{equation}\label{critpoint}
y=y_c = \frac{n-1}{n-2}\left[\frac{(n-2)c_n}{n \beta_n}\right]^{\frac{n}{2n-2}},\quad n>2,
\end{equation}
where the first derivative of the rate function $f_n(y)$ exhibits a discontinuity.

A question naturally emerges regarding the extent to which these anomalies and DPT are universal. In particular, can they arise if the stationary process $x(t)$ is non-Markovian, so that the Feynman-Kac method -- and, as a result, the DV theory -- do not apply?
A first step in addressing this question was taken in Ref. \cite{Meerson2019} which considered (in general non-Markovian) Gaussian stationary processes with short correlations and applied the optimal fluctuation method (OFM) to study the large deviations of $A_n$.  For $n>2$, the large deviations of  $A_n$ were found to exhibit the same  scaling behavior, $-\ln P(A_n,T) \sim A_n^{2/n}$, independent of $T$, as their counterparts for the OU process. This result showed that the breakdown of the simple large-deviations principle~(\ref{LDP}), previously observed in the OU process  \cite{NT2018}, also occurs for short-correlated Gaussian processes \cite{Meerson2019}.

Here we considerably extend the earlier work and study anomalous scaling and the ensuing DPT in the full probability distribution $P(A_n,T\to \infty)$ of stationary Gaussian processes: both short- and long-correlated. To remind the reader, a centered stationary Gaussian process  $x(t)$ is fully determined by its covariance
\begin{equation}
\kappa(s) = \langle x(t)x(t+s) \rangle.
\end{equation}
As we will see,  for \emph{short-correlated} Gaussian processes -- that is, when $\kappa(t)$ decays sufficiently fast at $|t|\to \infty$ -- the typical fluctuations of $A_n$ are normally distributed with a variance growing linearly with $T$.  This fact, combined with the $A^{2/n}$ scaling of the action for large $A_n$ \cite{Meerson2019}, implies that, for such processes, the probability distribution of $A_n$ for $n>2$ obeys an anomalous scaling of the type~(\ref{anomLDP}), with the same exponents $\mu$ and $\nu$ [given by Eq.~(\ref{munu})] as for the OU process. Furthermore, here too the rate function $f_n(z)$  is described by Eqs. (\ref{f}) and (\ref{F}) (with coefficients determined by the covariance of the particular Gaussian process), and it exhibits a first-order DPT of the same type.  

We will also show here that a similar first-order DPT can exist, under certain conditions, for \emph{long-correlated} Gaussian processes. 
In this case, however, the exponents $\mu$ and $\nu$ are determined not only by $n$, but also by the long-time power-law tail of the covariance. 

Furthermore, we show that, when the correlations are too long-ranged, the DPT disappears, giving way to a smooth crossover between the regions of typical, Gaussian fluctuations and large deviations. Overall, we obtain a phase diagram which shows regions of different behavior of the anomalous exponents $\mu$ and $\nu$ [see Eq.~(\ref{anomLDP})] and the presence or absence of the DPT in the rate function $f_n(y)$. Our general results are illustrated on several particular examples of stationary Gaussian processes. These include the fractional Ornstein-Uhlenbeck process \cite{Cheredito2003,Kaarakka2015}, where by varying the Hurst exponent $H$, one can observe  all the three regimes of anomalous behavior described above.  Finally, we corroborate our analytical predictions numerically by using replica-exchange Wang–Landau simulations, which allow us to measure probabilities as small as $10^{-200}$. Our numerics  includes the important case $H = 1/2$ (the standard OU process), for which the theoretical predictions (\ref{anomLDP})-(\ref{critpoint}) \cite{Smith2022} have so far not been tested numerically.

The layout of the remainder of the paper is the following. In Sec. \ref{short} we present our results for the short-correlated systems. The long-correlated systems are dealt with in Sec. \ref{Sec3}. Our large-deviation simulations and verification of our theoretical predictions  are described in Sec. \ref{simulations}. We summarize and briefly discuss our results in Sec. \ref{discussion}. 

\section{Short-correlated Gaussian processes}
\label{short}

\subsection{Typical fluctuations of $A_n$}
\label{Sec2a}

The typical fluctuations of $A_n$ around the mean follow the Gaussian distribution,
\begin{equation}
\label{typicalgauss}
    P(A_n,T) \simeq P_{\text{Gauss}}(A_n,T) = \frac{1}{\sqrt{2\pi \text{Var}_n}}\exp\left[-\frac{(A_n-\langle A_n \rangle)^2}{2 \text{Var}_n}\right],
\end{equation}
where $\text{Var}_n$ is the variance of $A_n$, and 
\begin{equation}
\label{meanA}
    \langle A_n \rangle =  \int\limits_0^T \langle x^n(t) \rangle dt  = T\frac{[1 + (-1)^n] [2\kappa(0)]^{n/2} \Gamma(\frac{n+1}{2})}{2\sqrt{\pi}}.
\end{equation}
Obviously $\langle A_n \rangle$ is nonzero only for even $n$. The second moment of $A_n$ can be written as
\begin{equation}
\label{varA}
    \langle A_n^2\rangle = \int\limits_0^T \int\limits_0^T \langle x^n(t)x^n(s) \rangle dt \,ds.
\end{equation}
The covariances $\langle x^n(t)x^n(s) \rangle$ can be evaluated either with the help of the bivariate normal distribution or, more conveniently, via
Isserlis-Wick theorem  \cite{AdlerTaylor}.  For any stationary Gaussian process and arbitrary $n=1,2,...$, Isserlis-Wick's theorem reduces this covariance to a finite sum over powers of pairwise covariances:
\begin{eqnarray}\label{kappa_n}
\langle x^n(t)x^n(s) \rangle =\sum\limits_{m=0}^{\left[\frac{n}{2}\right]} \frac{(n!)^2}{(n-2m)! (2^m m!)^2}\kappa^{2m}(0)\kappa^{n-2m}(t-s),
\end{eqnarray}
where $[z]$ is the integer part of $z$.  Substituting Eq.~(\ref{kappa_n}) into Eq.~(\ref{varA}) and subtracting the squared mean (\ref{meanA}), we obtain a general expression for the variance of $A_n$ (we denote it as $\text{Var}_n$) for an arbitrary stationary Gaussian process and for any $n$:
\begin{eqnarray}\label{var_An}
\text{Var}_n = \sum\limits_{m=0}^{\left[\frac{n-1}{2}\right]} \frac{(n!)^2}{(n-2m)! (2^m m!)^2}\kappa^{2m}(0)\int\limits_0^T \int\limits_0^T\kappa^{n-2m}(t-s)dt \,ds.
\end{eqnarray}

Crucially, for short-correlated processes the $T\to \infty$ asymptotic of $\text{Var}_n$ is proportional to $T$, because in this case
\begin{equation}
\int\limits_0^T \int\limits_0^T\kappa^{n-2m}(t-s) \,dt \,ds \simeq T \int_{-\infty}^{\infty} \kappa^{n-2m}(z) dz\,.
\end{equation}
As a result, the coefficient $\beta_n$ in Eq.~(\ref{betan}) is $T$-independent, which is a pre-requisite for the validity of Eqs.~(\ref{f}) and~(\ref{F}).  Let us illustrate Eq.~(\ref{var_An}) on two particular examples of Gaussian processes. 

\subsubsection{OU covariance}

The OU process is the only process (up to a simple rescaling) that is both Gaussian, and Markovian. 
By rescaling $t$ and $x$, we can choose its covariance to be parameter-free: $\kappa(t)=(1/2) \exp(-|t|)$.
Using Eq.~(\ref{var_An}),  
we can  calculate the $T\gg 1$ asympototic of the variance of $A_n$ for arbitrary $n$. The result, 
\begin{equation}
\label{OUvarn}
\text{Var}_n^{\text{OU}} \simeq T \sum_{m=0}^{\left[\frac{n-1}{2}\right]} \frac{(n!)^{2}}{(n-2m)(n-2m)! \,2^{n+2m-1} (m!)^{2}} ,
\end{equation}
is equivalent to Eq.~(14) of Ref. \cite{Bassanoni}. In particular, for $n=3,4,5$ and $6$  Eq.~(\ref{OUvarn})  yields
\begin{equation}
    \begin{cases}
       \text{Var}_3^{\text{OU}} \simeq \frac{11}{4}T, \quad \beta_3= \frac{2}{11}, \\
       \text{Var}_4^{\text{OU}} \simeq \frac{21}{4}T, \quad \beta_4= \frac{2}{21},\\
       \text{Var}_5^{\text{OU}} \simeq \frac{449}{16}T,\quad \beta_5 = \frac{8}{449},\\
       \text{Var}_6^{\text{OU}} \simeq \frac{3495}{32}T,\quad \beta_6 = \frac{16}{3495}.
    \end{cases}
\end{equation}
where we used Eq.~(\ref{betan}) to determine the coefficients $\beta_n$. For $n=3$ and $4$ these results coincide with those
obtained perturbatively within the framework of the DV theory \cite{Smith2022}. Equation~(\ref{OUvarn}) allows one to easily obtain $\beta_n$ for arbitrarily large $n$,
where the perturbative quantum-mechanical calculations \cite{Smith2022}  require a bit more technical effort.

\subsubsection{Gaussian covariance}
\label{gausscov}

Our second example deals with a centered stationary Gaussian process with a Gaussian covariance. Again, by rescaling $t$ and $x$ we can choose the covariance to be parameter-free: $\kappa(t)= \exp(-t^2)$. Plugging this covariance into Eq.~(\ref{var_An}),  
we obtain the $T\gg 1$ asympototic of the variance of $A_n$:
\begin{equation}
\label{Gaussvarn}
\text{Var}_n^{\text{Gauss}} \simeq \sqrt{\pi}\, T \sum_{m=0}^{\left[\frac{n-1}{2}\right]} \frac{(n!)^{2}}{\sqrt{n-2m}(n-2m)! \,2^{2m} (m!)^{2}}.
\end{equation}

For $n=3,4,5$ and $6$ Eqs.~(\ref{Gaussvarn}) and (\ref{betan}) yield
\begin{equation}\label{typical_gauss}
    \begin{cases}
       \text{Var}_3^{\text{Gauss}} \simeq (9 + 2\sqrt{3}) \sqrt{\pi} T, \quad \beta_3= \frac{1}{2 \left(9+2 \sqrt{3}\right) \sqrt{\pi}}, \\
       \text{Var}_4^{\text{Gauss}} \simeq 12 \left(1+3 \sqrt{2}\right)\sqrt{\pi} T, \quad \beta_4= \frac{1}{24 \left(1+3 \sqrt{2}\right) \sqrt{\pi}},\\
       \text{Var}_5^{\text{Gauss}} \simeq \left(225+200 \sqrt{3}+24 \sqrt{5}\right) \sqrt{\pi } T,\quad \beta_5 = \frac{1}{2 \left(225+200 \sqrt{3}+24 \sqrt{5}\right) \sqrt{\pi}},\\
       \text{Var}_6^{\text{Gauss}} \simeq15 \left(180+135 \sqrt{2}+8 \sqrt{6}\right) \sqrt{\pi} T,\quad \beta_6 = \frac{1}{30 \left(180+135 \sqrt{2}+8 \sqrt{6}\right) \sqrt{\pi}}.
    \end{cases}
\end{equation}
Again, these calculations are quite straightforward.

\subsection{Large deviations of $A_n$}
\label{sec2b}

The $A_n\to \infty$ tail of the probability distribution  $P(A_n,T)\sim \exp[-S(A_n,T)]$ is dominated by a single \emph{optimal} path, that is the most likely realization of the process $x(t)$ conditioned on a specified value of $A_n$ \cite{NT2018,Meerson2019}. The optimal path is the minimizer of
the Gaussian action $S[x(t)]$ along paths conditioned on the specified (large) value of $A_n$. 
For completeness, here we briefly recap, with slight changes, this calculation \cite{Meerson2019}. 

For a stationary Gaussian process the modified action functional to be minimized is the following:
\begin{eqnarray}
     S_{\lambda}[x(t)]=\frac{1}{2}\int\limits_{-\infty}^{\infty}dt \int\limits_{-\infty}^{\infty}dt' K(t -t')x(t)x(t')- \lambda\int\limits_{-\infty}^{\infty}dt x^n(t) \big[\theta(t)-\theta(t-T)\big]. \label{Act}
\end{eqnarray}
The  first term in Eq.~(\ref{Act}) determines (up to a pre-exponential factor) the statistical weight $P[x(t)]\sim \exp(-S[x(t)])$ of a given realization $x(t)$ of the process, and $K(\dots)$ is the inverse kernel of the stationary Gaussian process,  defined by the relation
\begin{equation}\label{inversek}
\int\limits_{-\infty}^{\infty}K(\tau)\kappa(t-\tau)d\tau=\delta(t).
\end{equation}
The second term in Eq.~(\ref{Act}) conditions the process on a large area $A_n$, see Eq.~(\ref{An}), with $\lambda$ playing the role of a Lagrange multiplier, and $\theta(\ldots)$ denoting the Heaviside step function.

The linear variation of the modified action~(\ref{Act}) must vanish, which yields a nonlocal analog of the Euler-Lagrange equation for the optimal path $x(t)$:
\begin{equation}
    \int\limits_{-\infty}^{\infty}dt' K(t -t')x(t')=n\lambda x^{n-1}(t)  \big[\theta(t)-\theta(t-T)\big].
    \label{IntEq}
\end{equation}
Multiplying both sides of this equation by  $\kappa(\tau)$ and using Eq.~(\ref{inversek}), one arrives at the integral equation  \cite{Meerson2019}
\begin{equation}
    x(t)=n\lambda \int\limits_{0}^{T}dt' \kappa(t -t')x^{n-1}(t').
    \label{IntEqkappa}
\end{equation}
Using Eq.~(\ref{IntEq}), one can express the action $S(A_n,T)$ as follows \cite{Meerson2019}:
\begin{eqnarray}
    S(A_n,T)=\frac{1}{2}\int\limits_{-\infty}^{\infty}dt\, x(t)\int\limits_{-\infty}^{\infty}dt'\, K(t -t') x(t')=\frac{n\lambda}{2}\int\limits_{0}^{T}dt x^n(t)=\frac{n\lambda A_n}{2},
    \label{act-labmda}
\end{eqnarray}
where the Lagrange multiplier $\lambda=\lambda(A_n)$ is to be ultimately determined from Eq.~(\ref{An}). 

At $n>2$ the integral equation (\ref{IntEqkappa}) is nonlinear, and the ensuing possible multiplicity of solutions plays a crucial role. In particular, an instanton solution emerges, which is localized on the characteristic correlation time scale defined by the covariance function $\kappa(t)$.   At sufficiently large $T$ this localized-in-time instanton solution has a lesser action  than an almost constant, delocalized solution [which would lead to the standard large-deviation scaling (\ref{LDP})].  The Gaussian action on such an instanton is \cite{Meerson2019}:
\begin{equation}
     S_{\text{inst}}(A_n,T) =  c_n A_n^{2/n},\quad T\to \infty,
    \label{scaling-action}
\end{equation}
where the coefficients $c_n$ are determined by the covariance function $\kappa(t)$ and by $n$. Equation~(\ref{scaling-action}) has the same form as its counterpart for the OU process \cite{NT2018}, and it determines the second term in the anomalous rate function~(\ref{F}).

For the standard OU process, the coefficients $c_n$ were found, for all $n>2$, in Ref. \cite{NT2018} (see also Refs. \cite{Meerson2019} and \cite{Smith2022}):
\begin{equation}\label{cnOU}
   c^\text{OU}_n = \frac{n}{4} \left[\frac{2\sqrt{\pi} \,\Gamma(\frac{n}{n-2})}{(n-2) \, \Gamma\left(\frac{3n-2}{2n-4}\right)}\right]^{\frac{n-2}{n}}\,,\quad n>2\,.
\end{equation}

For the Gaussian covariance $\kappa(t) = \exp(-t^2)$, the coefficients $c_n$ were obtained in Ref.~\cite{Meerson2019}:
\begin{equation}\label{c_n_Gaussian}
    c^\text{Gauss}_n = \frac{(n-1)^\frac{n-1}{n}}{2 \pi^\frac{1}{n}[n(n-2)]^\frac{n-2}{2n}}, \quad n>2\,.
\end{equation}

\subsection{Mixed region and DPT}
\label{2c}

Equations~(\ref{typicalgauss}) and~(\ref{scaling-action}) -- for the typical fluctuations and large deviations, respectively -- alongside with the $T$-independence of the coefficients $\beta_n$ and $c_n$ provide all the ingredients for the validity of the arguments of Ref. \cite{Smith2022}. This leads to the anomalous scaling~(\ref{anomLDP}), to the large deviation function~(\ref{f}) and~(\ref{F}), and to the first-order DPT at the critical point (\ref{critpoint}). For completeness we will now briefly reproduce the arguments of Ref. \cite{Smith2022}  

A crucial realization of Ref. \cite{Smith2022} was that there is an intermediate region of \emph{moderately large} deviations  of $\Delta A_n = A_n - \langle A_n \rangle$.  In this \emph{mixed} region a deviation of the empirical observable $\Delta A_n$ can be viewed as the sum of two almost independent contributions arising on well-separated time scales. The first of them, $\Delta A_{n,\text{inst}}$, comes from an instanton, which is localized on a short time interval $\tau_{\text{inst}}\ll T$ [a characteristic time scale of the covariance $\kappa(t)$]. The second contribution, $\Delta A_n - \Delta A_{n,\text{inst}}$, comes from fluctuations spread uniformly over the remaining time $T-\tau_\text{inst}$, and it obeys the central-limit-type Gaussian statistics~(\ref{typicalgauss}). 

Since these two mechanisms act on different time scales and, at $T\to \infty$, interact only weakly, 
the probability of observing a specified $\Delta A_n$ can be expressed as a convolution of the two separate contributions \cite{Smith2022},
\begin{equation}\label{P_condensation}
P(A_n,T) \simeq \int_0^{\Delta A_n} P_{\text{inst}}(\Delta A_{n,\text{inst}},T)\, P_{\text{Gauss}}\big(\Delta A_n - \Delta A_{n,\text{inst}},T\big)\, d\Delta A_{n,\text{inst}},
\end{equation}
where the integration is over the instanton contributions. 
In the large-$T$ limit, 
this integral is dominated by the saddle-point of the effective joint action, describing the interplay between the typical and large fluctuations:
\begin{equation}\label{S_mixed}
    S_{\text{mixed}}(\Delta A_n,\Delta A_{n,\text{inst}},T) =\frac{\beta_n(\Delta A_n -\Delta A_{n,\text{inst}})^2}{T}+ c_n \Delta A_{n,\text{inst}}^{2/n}.
\end{equation}
The saddle point is located in the region where the two contributions are comparable, that is where $\Delta A_n \sim \Delta A_{n,\text{inst}} \sim T^{n/(2n-2)}$. (This implies that, as $T$ goes to infinity, one should keep $\Delta A_n/T^{n/(2n-2)}$ constant  \cite{Smith2022}).
The saddle-point evaluation yields Eqs.~(\ref{anomLDP})-(\ref{F}) for the probability distribution and Eq.~(\ref{critpoint}) for the critical point of the first-order DPT \cite{Smith2022}.

As one can see, the main findings of Ref. \cite{Smith2022}, obtained for stationary Markov  processes, can be directly extended to (in general non-Markovian) \emph{short-correlated} Gaussian processes. In Sec. \ref{simulations}  we compare these predictions with replica-exchange Wang–Landau simulations for the Gaussian covariance introduced in Sec. \ref{gausscov}.

\section{Long-correlated Gaussian processes}
\label{Sec3}

Now let us consider \emph{long-correlated} stationary Gaussian processes $x(t)$  whose covariance
exhibits a power-law tail,
\begin{equation}\label{tail}
\kappa(t\to\infty)\simeq B |t|^{-\alpha}\,,\quad \alpha>0\,.
\end{equation}
A  heavy tail corresponds to  $\alpha<1$. We start with the typical fluctuations of $\Delta A_n$. 

\subsection{Typical fluctuations}

Since the integrand in Eq.~(\ref{var_An}) depends only on the difference $t-s$, we can rewrite the double integral as a single integral:
\begin{eqnarray}\label{var_An_modified}
\text{Var}_n = \sum\limits_{m=0}^{\left[\frac{n-1}{2}\right]} \frac{(n!)^2}{(n-2m)! (2^m m!)^2}\kappa^{2m}(0)\int\limits_{-T}^T (T-|t|)\kappa^{n-2m}(t)dt.
\end{eqnarray}
Each term of the sum with index $p=n-2m$ involves the integral
\begin{equation}
I_p(T) = \int_{-T}^{T}(T-|t|)\kappa^p(t)dt.
\end{equation}
Using the covariance tail asymptotic~(\ref{tail}), one can identify three distinct regimes of the large-$T$ scaling behavior of $I_p(T)$, depending on the integrability of the $p$-th power of the covariance, $\kappa^p(t)$, at $T\to\infty$ : 
\begin{equation}
    I_p(T)\sim\begin{cases}
        T^{E_p(\alpha)},&  \text{with $E_p(\alpha)=2-p\alpha$, for $p\alpha<1$} ,\\
        T,&  \text{for $p\alpha>1$} ,\\
        T\ln T,&  \text{for $p\alpha=1$} .\\
    \end{cases}
\end{equation}
Since $\alpha>0$, the exponent $E_p(\alpha)$ decreases with $p$. Therefore, the dominating contribution to the variance comes from the  term with the largest $E_p(\alpha)$ or, equivalently, with the lowest possible index $p_\text{min}$  in Eq.~(\ref{var_An_modified}), which is
\begin{equation}\label{pmin}
    p_\text{min} = \begin{cases}
    1,&  \text{for odd $n$} ,\\2,&  \text{for even $n$}.\end{cases}
\end{equation}
This yields a threshold value $\alpha=\alpha_c$ separating the ``normal" regime $\alpha>\alpha_c$, where the variance grows linearly with $T$, and the regime of an anomalous scaling of the variance with $T$ at $\alpha<\alpha_c$:
\begin{equation}\label{alpha_c}
\alpha_c=\frac{1}{p_{\min}} = 
    \begin{cases}
    1, &  \text{for odd $n$} ,\\
    1/2,&  \text{for even $n$}.
    \end{cases}
\end{equation}
Altogether, we have at $T\to \infty$:
\begin{equation}\label{Var_anomalous}
\text{Var}_n\sim \begin{cases}
T,&  \text{for $\alpha\ge \alpha_c$} ,\\
T^E,&  \text{for $\alpha<\alpha_c$} ,\\
\end{cases}
\quad\text{where}\quad
\alpha_c=\begin{cases}1,&  \text{for odd $n$},\\ 1/2,& \text{for even $n$},\end{cases}\quad \text{and}\quad
E=\begin{cases}2-\alpha,&  \text{for odd $n$},\\ 2-2\alpha,& \text{for even $n$}.\end{cases}
\end{equation}
Using  Eq.~(\ref{betan}), we determine the scaling behavior of $\beta_n$ with $T$:
\begin{equation}\label{betan_longranged}
\quad \beta_n \sim \begin{cases}
1,&  \text{for $\alpha\ge \alpha_c$} ,\\
T^{\alpha-1},&  \text{for $\alpha<\alpha_c$ and odd $n$} ,\\
T^{2\alpha-1},&  \text{for $\alpha<\alpha_c$ and even $n$}.\\
\end{cases}
\end{equation}
Importantly, for $\alpha<\alpha_c$, the coefficients $\beta_n$ depend on $T$, which reflects a faster-than-linear scaling behavior 
of the variance with $T$ in this regime.

\subsection{Large deviations}

Large deviations of $A_n$ for stationary Gaussian processes with long-range correlations, described by Eq.~(\ref{tail}), have been recently studied in Ref.~\cite{VM2025} in the particular case of the fractional
Ornstein-Uhlenbeck process. Similarly to the short-correlated case, a key factor in determining the large-deviation tail of 
$P(A_n,T)$ is a competition between localized (instanton-type) and delocalized solutions. The $T$-independent characteristic duration of the instanton solution, $\tau_\text{inst}\ll T$, is
determined by the intrinsic time scale of the covariance $\kappa(t)$, and the presence of such a time scale is necessary for the existence of instanton. Further, there is a critical value $\alpha=\alpha(n)$ at which a transition occurs between localized and delocalized solutions of Eq.~(\ref{IntEqkappa}), and correspondingly, between distinct scaling behaviors of the large-deviation action~(\ref{scaling-action}).

The action corresponding to a localized solution of Eq.~(\ref{IntEqkappa}) exhibits the same scaling behavior with $T$, see Eq.~(\ref{scaling-action}), as that for the short-correlated case. To estimate the action corresponding to a \emph{delocalized} solution, one can assume that it is almost constant \cite{VM2025}.  Then, using the large-$t$ asymptotic of the covariance~(\ref{tail})  in Eq.~(\ref{IntEqkappa}) and integrating both sides of the equation over $(0,T)$, we obtain the following order-of-magnitude relation:
\begin{equation}
    \text{const}^{(n-2)} \sim T \left( n\lambda\int\limits_{0}^{T} \int\limits_{0}^{T} \kappa(t - s) \, dt \, ds \right)^{-1} \sim \lambda^{-1}B^{-1} T^{\alpha-1}, \quad T\to \infty.
\end{equation}
Expressing the corresponding Lagrange multiplier $\lambda$ through $A_n$ and using Eqs.~(\ref{Act}) and (\ref{IntEq}), we find that the action, associated with an  (almost constant)  delocalized optimal path, scales as
\begin{eqnarray}\label{action_const}
    S_\text{const}\sim  T^{\alpha-2/n} A_n^{2/n}.
\end{eqnarray}
This action decreases with $T$ for 
\begin{equation}\label{alpha_cond}
    0<\alpha<\alpha_* \equiv 2/n,
\end{equation} 
making the delocalized solution favorable as  $T \to \infty$. For $\alpha\geq \alpha_*$, a localized solution yields a $T$-independent action~(\ref{scaling-action}) which is therefore optimal. Overall, the scaling behavior of the large-deviation action for long-correlated stationary Gaussian processes is the following:
\begin{equation}
    S(A_n,T)=c_n A_n^{2/n},\quad\text{with}\quad c_n\sim\begin{cases}
        1,\qquad & \alpha>\alpha_*,\\
        T^{\alpha-2/n}, \qquad & \alpha<\alpha_*.
    \end{cases} 
    \label{cn_scaling}
\end{equation}

\subsection{Mixed region}

Even when one or both of the coefficients $\beta_n$ and $c_n$  in Eq.~(\ref{S_mixed}) are $T$-dependent, we can still predict the scaling behavior $\Delta A_n \sim T^\nu$ by balancing in Eq.~(\ref{S_mixed})  the instanton contribution $c_n A_n^{2/n}$ with the contribution from  the typical, Gaussian fluctuations, $\beta_n A_n^2/T$:
\begin{equation}\label{nu_scaling}
    c_n T^{2\nu/n}\sim \frac{\beta_n}{T}T^{2\nu}.
\end{equation}
Combining this relation with the conditions~(\ref{betan_longranged}) and (\ref{cn_scaling}),  we identify three distinct scaling regimes for the distribution  $P(A_n,T)\sim \exp\left[-T^\mu f_n\left(\Delta A_n T^{-\nu}\right)\right]$.

In the regime $\alpha> \alpha_c$ (that is, for correlation tails which are not heavy or not too heavy, for odd and even $n$, respectively), both coefficients $\beta_n$ and $c_n$ are independent of $T$.   In this regime, the probability destribution exhibits the anomalous scaling  and the first-order DPT characteristic of the short-correlated processes, as described by Eqs.~(\ref{anomLDP})-(\ref{F}) for the probability distribution and Eq.~(\ref{critpoint}) for the critical point \cite{Smith2022}. 

In the regime of $\alpha_*\le \alpha<\alpha_c$, the long-range correlations modify the scaling of the typical fluctuations of $A_n$, so that the coefficients $\beta_n$ depend on $T$. The large deviations, however, are still governed by a localized instanton-like optimal path, so that $c_n$ is $T$-independent. Therefore, as $T\to \infty$, the first-order DPT should persists, but it shifts toward larger $A_n$. Here the anomalous scaling of the probability distribution is described by Eq.~(\ref{anomLDP}) with the exponents $\mu$ and $\nu$ which depend both on $n$ and on $\alpha$:
\begin{equation}
   \mu =\begin{cases}\frac{2(2-\alpha)}{2n-2},&  \text{for odd $n$},\\ \frac{2(2-2\alpha)}{2n-2},& \text{for even $n$},\end{cases}\,\,\,
      \,\,\,\nu=\begin{cases}\frac{n(2-\alpha)}{2n-2},&  \text{for odd $n$},\\ \frac{n(2-2\alpha)}{2n-2},& \text{for even $n$}.\end{cases}
\end{equation}

In the regime of very long correlations, $\alpha<\alpha_*$, both $\beta_n$ and $c_n$ are $T$-dependent. Here  large but almost uniform deviations in time have a lower action cost than localized instantons. Consequently, the convolution construction~(\ref{P_condensation}) does not apply, and a DPT is absent.  The system therefore should exhibit a smooth crossover between the typical and large fluctuations. The crossover is expected to occur at $\Delta A_n \sim T^\nu$, where  the Gaussian action of the typical fluctuations is comparable with the action corresponding to a delocalized optimal path which determines the large deviations of $A_n$, and we obtain Eq.~(\ref{anomLDP}) with 
\begin{equation}
    \,\,\, \mu =\begin{cases}\alpha,&  \text{for odd $n$},\\ \frac{n-2}{n-1}\alpha,& \text{for even $n$}, \end{cases}
   \,\,\,\,\,\,\nu=\begin{cases}1,&  \text{for odd $n$},\\ 1-\frac{n\alpha}{2n-2},& \text{for even $n$},\end{cases}
\end{equation}
and $f_n(y)\sim \mathcal{O}(1)$ is a smooth function of $y$. 

\begin{figure}[ht]
\centering
\includegraphics[clip,width=0.35\textwidth]{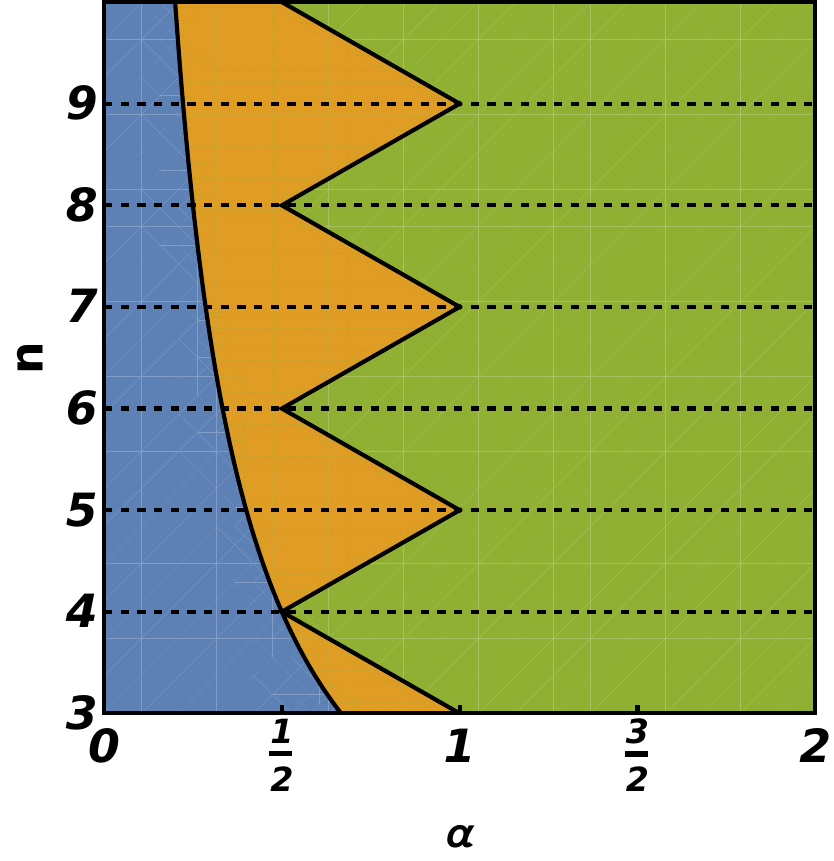}
\caption{Phase diagram of the anomalous scaling behavior of the distribution $P(A_n,T)\simeq \exp\left[-T^\mu f_n\left(\Delta A_n T^{-\nu}\right)\right]$ for $n>2$ on the $(\alpha, n)$ plane. Different colors correspond to regions with a different behavior of the scaling exponents $\mu$ and $\nu$. The green region represents the effectively short-correlated regime $\alpha> \alpha_c$, where the rate function $f_n(y)$ exhibits a first-order DPT  identified in Ref. \cite{Smith2022}. The yellow region corresponds to the moderately correlated regime with $\alpha_*\le \alpha<\alpha_c$, which also displays a first-order DPT, but with a different behavior of $\mu$ and $\nu$. The blue region corresponds to the long-correlated regime $\alpha<\alpha_*$, where the rate function $f_n(y)$ is analytic. For illustrative purposes we have extended the phase diagram to non-integer $n$, but the results are applicable only for integer $n>2$, as indicated by the  dashed lines.}
\label{fig_PD_alpha_n}
\end{figure}

The resulting phase diagram on the  $(\alpha, n)$ plane, shown in Fig.~\ref{fig_PD_alpha_n}, describes  three distinct regimes of the anomalous scaling of the distribution $P(A_n,T)\simeq \exp\left[-T^\mu f_n\left(\Delta A_n T^{-\nu}\right)\right]$. The corresponding values of the exponents $\mu$ and $\nu$ are presented in Table~\ref{tab_1}.

\begin{table}[h!]
\centering
\caption{Anomalous scaling behavior of  $P(A_n,T)\simeq \exp\left[-T^\mu f_n\left(\Delta A_n T^{-\nu}\right)\right]$ for stationary Gaussian processes with long-ranged correlations, $\kappa(t\to\infty)\simeq B |t|^{-\alpha}$ at $|t| \to \infty$. The two critical exponents,  $\alpha_c$ from Eq.~(\ref{alpha_c}) and $\alpha_*=2/n$, separate three distinct scaling regimes.}
\vspace{6pt}
\renewcommand{\arraystretch}{2.0}
\begin{tabular}{|c|c|c|}
\hline
 $\alpha$ &  $\mu$ and $\nu$ & Behavior of \textbf{$f_n(y)$} \\ \hline

$\alpha > \alpha_c$ &
$\displaystyle \mu = \frac{2}{2n-2}, \quad
\nu = \frac{n}{2n-2}$ &
nonanalytic  \\ \hline

$\alpha_*\le \alpha < \alpha_c$ &
\begin{tabular}{@{}c@{}}
Odd $n$: $\displaystyle
\mu = \frac{2(2-\alpha)}{2n-2}, \;
\nu = \frac{n(2-\alpha)}{2n-2}$ \\[4pt]
Even $n$: $\displaystyle
\mu = -\frac{2(2-2\alpha)}{2n-2}, \;
\nu = -\frac{n(2-2\alpha)}{2n-2}$
\end{tabular} &
nonanalytic  \\ \hline

$0<\alpha < \alpha_*$ &
\begin{tabular}{@{}c@{}}
Odd $n$: $\displaystyle
\mu = \alpha, \;
\nu = 1$ \\[4pt]
Even $n$: $\displaystyle
\mu = \frac{n-2}{n-1}\alpha, \;
\nu = 1 - \frac{n\alpha}{2(n-1)}$
\end{tabular} &
analytic  \\ \hline

\end{tabular}
\label{tab_1}
\end{table}

Interestingly, for $n=4$ the critical exponents $\alpha_*$ and $\alpha_c$ coincide: $\alpha_*=\alpha_c=1/2$. As a result, there is no yellow region in this case.

\subsection{fOU process}

A representative example of a stationary Gaussian process  with a power-law asymptotic behavior of the correlation 
is provided by the fractional Ornstein–Uhlenbeck (fOU) process \cite{Cheredito2003,Kaarakka2015}, which has recently attracted a renewed attention in the context of large deviations \cite{MS2024,VM2025}. The fOU process is governed by the linear overdamped Langevin equation :
\begin{equation}
    \dot{x}(t)+\gamma x=\sqrt{2 D}\xi(t),
    \label{LangevinfOU}
\end{equation}
where $D>0$ is the coefficient of fractional diffusion, and $\xi(t)$ is the fractional Gaussian noise: a centered stationary Gaussian process with the covariance
\begin{equation}
    c(\tau)=\langle \xi(t+\tau)\xi(t) \rangle=\frac{d}{d\tau}\big(H \vert \tau\vert^{2H-1}\text{sgn}(\tau)\big)=\frac{1}{2}\frac{d^2}{d\tau^2}\vert \tau \vert^{2H}\,,
\label{kappa}
\end{equation}
where $0<H<1$ is the Hurst exponent. The covariance $\kappa(\tau)=\langle x(t+\tau) x(t)\rangle$ of the stationary fOU process is given by
\begin{equation}
    \kappa(\tau)=\frac{1}{2\pi}\int\limits_{-\infty}^{\infty}
    \kappa_{\omega} e^{-i\omega \tau}d\omega=\frac{D  \Gamma (2 H+1)}{\gamma ^{2 H}} \left[\cosh (\gamma  \tau)-\frac{ |\gamma \tau| ^{2 H} }{\Gamma (2 H+1)}{}_1F_2\left(1;H+\frac{1}{2},H+1;\frac{\tau^2 \gamma ^2}{4}\right)\right],
    \label{cov}
\end{equation}
where $_1F_2(\ldots)$ is the hypergeometric function \cite{Wolfram}. 
At long times this covariance exhibits a power-law tail
\begin{equation}
\kappa(|t| \to \infty) \simeq \frac{2DH(2H-1)}{\gamma^{2}} |t|^{-(2-2H)},
\end{equation}
which corresponds to $\alpha=2-2H$ in Eq.~(\ref{tail}). With this $\alpha$, Eqs.~(\ref{alpha_c}) and ~(\ref{alpha_cond}) yield the critical values $H_c$ and  $H_*$ which separate the regimes of the ordinary and the anomalous \emph{typical} fluctuations, and of the localized and delocalized optimal paths, respectively:
\begin{equation}
H_c=\begin{cases}1/2,&  \text{for odd $n$},\\ 3/4,& \text{for even $n$},\end{cases}\quad \text{and}\quad
H_*=1-1/n.
\end{equation}

For general $H$ the integral, which appears in Eq.~(\ref{var_An_modified}) for the variance of $A_n$, can be evaluated numerically.  The calculation of the coefficients $c_n$, which describe the large deviations, can be done via an iterative numerical solution of the nonlinear integral equation~(\ref{IntEq}), see Sec.~\ref{Sec4a}. As an example, Table~\ref{tab_num} presents the numerically obtained $T$-independent coefficients $\beta_3$ and $c_3$, and the predicted values of the DPT point $y_c$, for $n=3$ and $H=1/4$. (For convenience, here we are using the units where $D=\frac{\gamma^{2H}}{\Gamma(2H+1)}$ and $\gamma=1$.)
The $T$-dependent coefficients ($\beta_3$ for $H=3/5$, and  both $\beta_3$ and $c_3$ for $H=3/4$) were obtained by fitting numerically the computed values of $c_3$ and $\beta_3$ by the power-law functions $\beta_3=\tilde{\beta}_3T^{1-2H}$ and $c_3=\tilde{c}_3 T^{1-2H-2/n}$, respectively,  over a range of $T$ from $10^2$ to $4\cdot 10^3$. 

\begin{table}[h!]
\centering
\caption{Numerically computed coefficients $\beta_3$ and $c_3$ for the fOU process with different $H$. }
\vspace{6pt}
\renewcommand{\arraystretch}{2.0}
\begin{tabular}{|c|c|c|c|}
\hline
$H$ & \textbf{$\beta_3$} & \textbf{$c_3$} & \textbf{$y_3$} \\ \hline

$ 1/4$ &
$ 0.095$ &
$0.751$ &
$4.134$ \\ \hline

$ 3/5$ &
$0.028T^{-1/5}$ &
$0.363$ &
$5.999$ \\ \hline

$ 3/4$ &
$0.036T^{-1/2}$ &
$0.332T^{-1/6}$ &
$-$ \\ \hline

\end{tabular}
\label{tab_num}
\end{table}

\begin{figure}[ht]
\centering
\includegraphics[clip,width=0.35\textwidth]{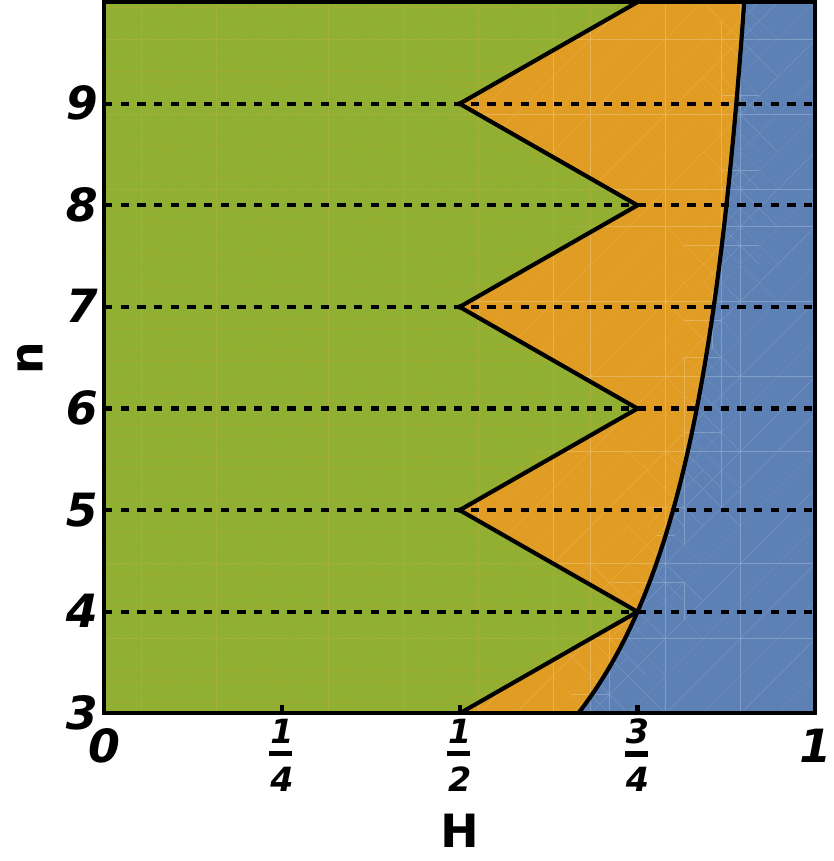}
\caption{Phase diagram of the anomalous scaling behavior of the distribution $P(A_n,T)$ for $n>2$ on the $(H, n)$ plane. Different colors correspond to regions with a different behavior of the scaling exponents $\mu$ and $\nu$. The green region represents the short-correlated regime $H\le H_c$, where the rate function exhibits a first-order DPT. The yellow region corresponds to the moderately correlated regime with $H_*\ge H>H_c$, which also displays a first-order DPT, but with a different behavior of $\mu$ and $\nu$. The blue region corresponds to the long-correlated regime $H>H_*$, where the DPT disappears. For illustrative purposes, we have extended the phase diagram to non-integer $n$, but the results are applicable only for integer $n>2$, as indicated by the  dashed lines.}
\label{fig_PD_H_n}
\end{figure}

Figure ~\ref{fig_PD_H_n} shows the phase diagram of the fOU process in the $(H,n)$ plane. The critical Hurst exponents $H_c$ and $H_*$ define three distinct regions in the $(H,n)$ plane. These regions determine the scaling exponents $\mu$ and $\nu$ of the asymptotic distribution (\ref{anomLDP}) in the large $T$ limit, see Table \ref{tab_2}. 
To remind the reader, for $n=4$ the yellow region is absent.

\begin{table}[h!]
\centering
\caption{Fluctuation regimes and scaling behaviors for the fOU process. 
The critical exponents $H_c$ and $H_*=1-1/n$ separate three distinct regimes corresponding to different scaling and large-deviation behaviors of $P(A_n,T)$.}
\vspace{6pt}
\renewcommand{\arraystretch}{2.0}
\begin{tabular}{|c|c|c|}
\hline
$H$&  $\mu$ and $\nu$ & Behavior of $f_n(y)$ \\ \hline

$H \le H_c$ &
$\displaystyle \mu = \frac{2}{2n-2}, \quad
\nu = \frac{n}{2n-2}$ &
nonanalytic  \\ \hline

$H_c < H \le H_*$ &
\begin{tabular}{@{}c@{}}
Odd $n$: $\displaystyle
\mu = \frac{4H}{2n-2}, \;
\nu = \frac{2nH}{2n-2}$ \\[4pt]
Even $n$: $\displaystyle
\mu = \frac{4(2H-1)}{2n-2}, \;
\nu = \frac{2n(2H-1)}{2n-2}$
\end{tabular} &
nonanalytic  \\ \hline

$H > H_* $ &
\begin{tabular}{@{}c@{}}
Odd $n$: $\displaystyle
\mu = 2 - 2H, \;
\nu = 1$ \\[4pt]
Even $n$: $\displaystyle
\mu = \frac{n-2}{n-1}(2 - 2H), \;
\nu = 1 + \frac{n(2H+2)}{2(n-1)}$
\end{tabular} &
analytic  \\ \hline

\end{tabular}
\label{tab_2}
\end{table}


\section{Large Deviation Simulations}
\label{simulations}
\subsection{Replica-Exchange Wang-Landau Simulations}
\label{Sec4a}

The predicted DPT can be observed only at very large $T$ which are inaccessible to conventional Monte-Carlo simulations. Therefore, we employed the Replica-Exchange Wang-Landau (REWL) algorithm \cite{Vogel2013, Vogel2014,Vogel2018} to probe the probability density $P(A_n,T)$ and to sample discretized configurations of stationary Gaussian processes. The essence of the REWL approach lies in dividing the target space of $A_n$ into multiple smaller overlapping windows. Every window runs one or more replicas. Each replica carries out an ordinary Wang–Landau (WL) sampling process \cite{WL1,WL2} with its own histogram  and density of states  within the corresponding window. After a fixed number of Wang–Landau steps, replica exchanges between the overlapping windows are performed. The probability of accepting a replica exchange, $P(X\leftrightarrow Y)$, of configurations $X$ and $Y$ between the windows $i$ and $j$ is given by
\begin{equation}
    P(X\leftrightarrow Y) = \min \left[1,\frac{S_i(A_X)S_j(A_Y)}{S_i(A_Y)S_j(A_X)}\right],
\end{equation}
where $S_i(A_X)$ is the current estimator for the density of states, $S(A_X,T) = -\ln P(A_X,T)$, in the window $i$, for the value $A_X = \int\limits_0^T X^n(t)dt$ corresponding to its configuration $X$. Once the simulation is complete, i.e. when the slowest replica reaches the desired precision, the overlapping fragments of the density of states are merged by appropriately shifting adjacent segments \cite{Vogel2018}.

The WL simulations were performed by discretizing the continuous process $X(t)$ over a large but finite domain $(0,T_D]$. The discretization is represented as $\Vec{X} = ( X(\Delta t), \ldots, X(T_D))$, where $\Delta t$ is the lattice step, and $L \equiv T_D/\Delta t \gg 1$ is the duration of the discretized trajectory. The dynamical observable $A_n$  of the discretized process $\Vec{X}$ is approximated by the discrete sum
\begin{equation}
    A_n=\Delta t\sum\limits_{k=L/2-T/2\Delta t}^{L/2+T/2\Delta t} X^n(k\Delta t).
\end{equation} 
Since a direct sampling of Gaussian stationary processes is computationally expensive, we employed the Circulant Embedding Method (CEM) \cite{Chan1994,Dietrich1997}, which enables sample generation in $O(L \ln L)$ time.  For more details on the WL simulations of stationary Gaussian processes see Ref. \cite{VLM2024}, where the CEM and the WL algorithm were used to study the large deviations of the potential barrier height distribution of statistically homogeneous Gaussian disorder potentials.

An important issue in these simulations is a proper choice of the lattice step $\Delta t$ to resolve the instanton-like optimal paths of $X(t)$ and accurately measure the corresponding action. To address this issue, we 
solved numerically, by a straightforward iteration procedure, a discretized version of the integral equation~(\ref{IntEqkappa}) which describes the optimal paths in the limit of large $A_n$:
\begin{equation}
  y_i(m \Delta t)=\Delta t\sum\limits_{k=L/2-T/2\Delta t}^{L/2+T/2\Delta t} \kappa((m-k)\Delta t)y_{i-1}^{n-1}(k\Delta t),
    \label{yn_d}
\end{equation}
where $y(t)=(n\lambda)^{1/(n-2)} x(t)$ is the rescaled optimal path. Similarly to Ref.~\cite{VM2025}, we reduced the discretization step $\Delta t$ until the numerical solution $y(t)$ converged with a required accuracy. Then we calculated the action $S(A_n,T)$ of the optimal path, leading to a discrete approximation of $c_3$. Additional details of the numerical implementation of the iteration process are presented in the Appendix of Ref. \cite{VM2025}.

We checked that for $\Delta t < 1.0$ the discrete approximation of the coefficient $c_3$ for the Gaussian covariance $\kappa(t)=\exp(-t^2)$ closely matches the exact analytical value given in Eq.~(\ref{c_n_Gaussian}).  In the REWL simulations of this case we used $\Delta t = 0.5$, resulting in a relative discretization error of the order of $10^{-11}$.

The REWL simulations ran for several integration times: $T=500,\,1000,\,2000,\, 4000$, and $8000$. The duration of the continuous trajectories -- $T_D = 512,\, 1024,\, 2048,\, 4096$, and $8192$, respectively -- were chosen to be large enough to minimize the edge effects in the integration region while still be manageable for an effective sampling.  (The durations of the discretized trajectory were chosen to be powers of 2 to optimize the FFT efficiency.)

Finally, the parameters $D$ and $\gamma$ were chosen to be $D=\frac{\gamma^{2H}}{\Gamma(2H+1)}$ and $\gamma=1$, while the discretization step $\Delta t$ was kept the same as for the Gaussian-correlated process.  Following recommendations of Ref. \cite{Vogel2014}, the overlap between the windows for all replicas in the simulations was set to $75\%$.

\subsection{Theory versus simulations}

We now present the numerical results for $n=3$ for the Gaussian-correlated process, as an example of short-correlated Gaussian processes, and for the fOU process -- a process which interpolates between the short- and intermediately-correlated regimes, which exhibit the first-order DPTs, and the long-correlated regime with a smooth crossover between the Gaussian fluctuations and the large deviations.

\subsubsection{Gaussian covariance}

\begin{figure}[ht]
\centering
\includegraphics[clip,width=0.6\textwidth]{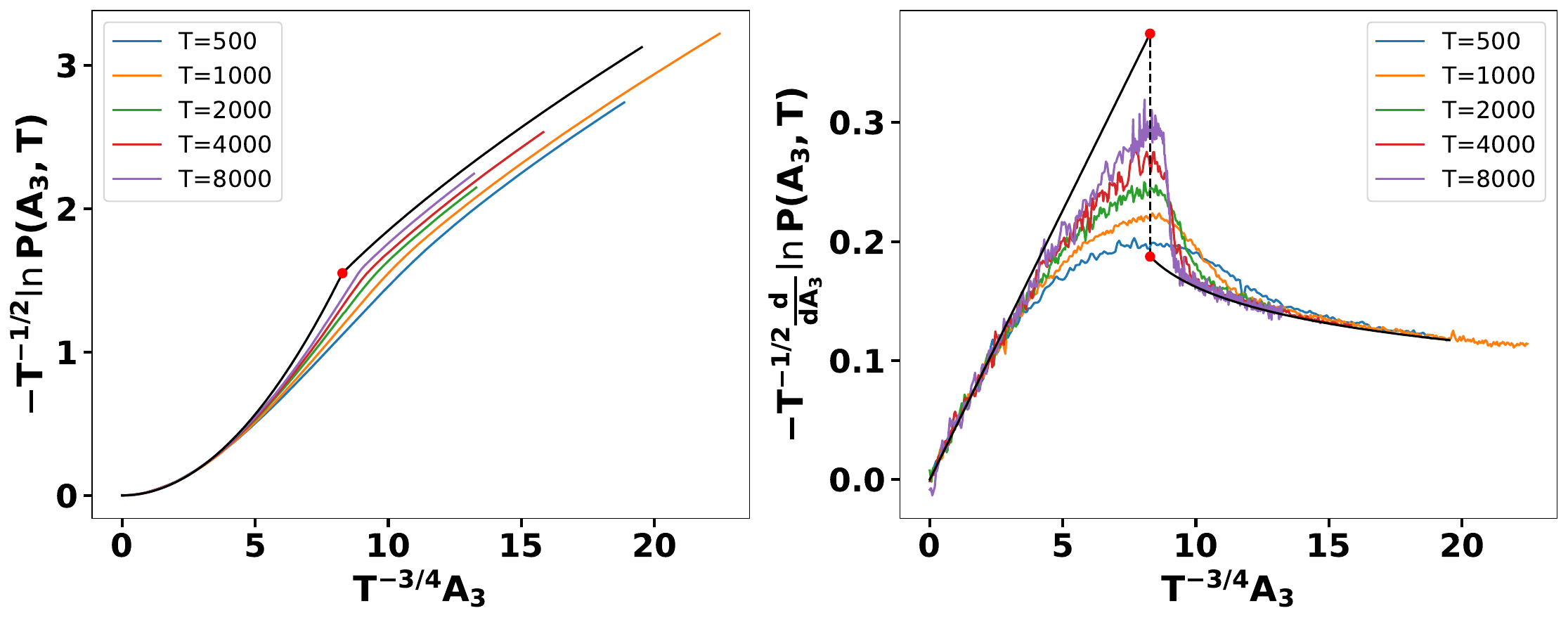}
\caption{The emerging first-order DPT for the Gaussian covariance. Left: the large-$T$ behavior of the rescaled action, $-T^{-1/2}\ln P(A_3,T)$, as measured in the REWL simulations for a set of values of $T\gg 1$ marked by different colors.  Right: the first derivative of the rescaled action with respect to $A_3$. The red dots mark the predicted critical point $y_c$ (left) and the finite discontinuity of the derivative (right).}
\label{fig_2}
\end{figure}

Combining Eqs.~(\ref{critpoint}), (\ref{typical_gauss}) and  (\ref{c_n_Gaussian}), obtained for the process with the Gaussian covariance $\kappa(t)= \exp(-t^2)$, we determine the critical point of the first-order DPT for $n=3$: $y_c\simeq 8.28$. Our simulation results for the rescaled action  $-T^{-1/2}\ln P(A_3, T)$ are shown in the left panel of Fig. \ref{fig_2} along with the theoretically predicted behavior, see Eqs.~(\ref{f}) and ~(\ref{F}). The black lines correspond to the Gaussian approximation $b_3 y^2$ for $y<y_c$ and to the mixed scenario for $y>y_c$, the red dots mark the predicted critical point $y_c$ and the finite discontinuity of the first derivative at $T\to \infty$.

\subsubsection{fOU process}

\begin{figure}[ht]
\centering
\includegraphics[clip,width=1.0\textwidth]{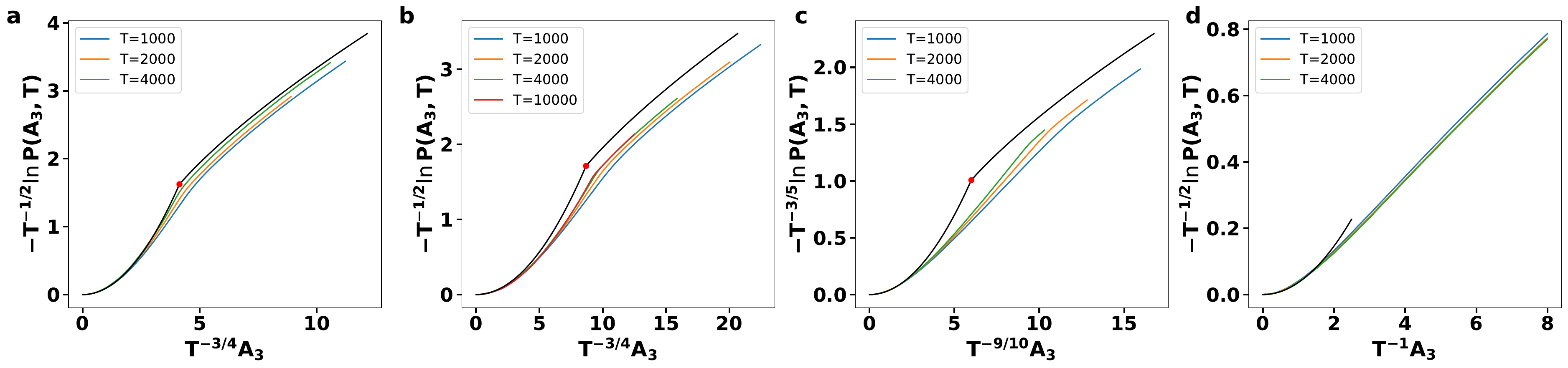}
\caption{The large-$T$ behavior of the rescaled action, $-T^{-\mu}\ln P(A_3,T)$, as measured in the REWL simulations of the fOU process for several values of $T$ (marked by different colors) and a set of Hurst exponents $H$: a) $H=1/4$; b) $H = 1/2$; c) $H=3/5$, and d) $H=3/4$. The red dots mark the predicted critical point $y_c$.}
\label{fig_RF}
\end{figure}

The simulation results for the rescaled action $-T^{-\mu}\ln P(A_3,T)$  of the fOU process for a set of Hurst exponents $H$ are shown in  Fig.~\ref{fig_RF}a-d along with the predicted behavior, see Table~\ref{tab_2}. For completeness, this set includes the important case $H = 1/2$ (the standard OU process), for which the theoretical predictions \cite{NT2018,Smith2022} have so far not been tested numerically.
The black lines correspond to Eqs.~(\ref{f}) and ~(\ref{F}) with the parameters $\beta_3$ and $c_3$ given in Table~\ref{tab_num}. In the short-correlated regime (Fig.~\ref{fig_RF}a,b), corresponding to $H = 1/4$ and  $H = 1/2$, the scaling exponents $(\mu, \nu)$ are universal and independent of the details of the covariance function.  For $H=3/5$ the covariance tail of the fOU process is sufficienly long-correlated to affect the typical fluctuation scaling, see Eq.~(\ref{var_An_modified})  and Table~\ref{tab_num}. Here the scaling behavior of the action $-\ln P(A_3,T)$ becomes $H$-dependent (see Table~\ref{tab_2} for $H_*\ge H>H_c$). The black lines in Fig.~\ref{fig_RF} represent the Gaussian approximation $\tilde{b}_3 y^2$ for $y<y_c$ and, in Figs. a-c,  the supercritical asymptotic $c_3 y^{2/3}$ for $y>y_c$, with the coefficients  $\tilde{b}_3$ and $c_3$ given in Table~\ref{tab_num}. Here $\tilde{\beta}_3=\beta_3T^{2H-1}$ is $T$-independent. In the strongly-correlated case $H=3/4$, presented in Fig.~\ref{fig_RF}d, the DPT disappears completely -- again, as predicted by the theory. Instead, a smooth crossover is observed between the typical fluctuations, described by the Gaussian asymptotic $\tilde{\beta_3}y^2$, and the large  deviations, described by the asymptotic $\tilde{c}_3 y^{2/3}$.

\begin{figure}[ht]
\centering
\includegraphics[clip,width=1.0\textwidth]{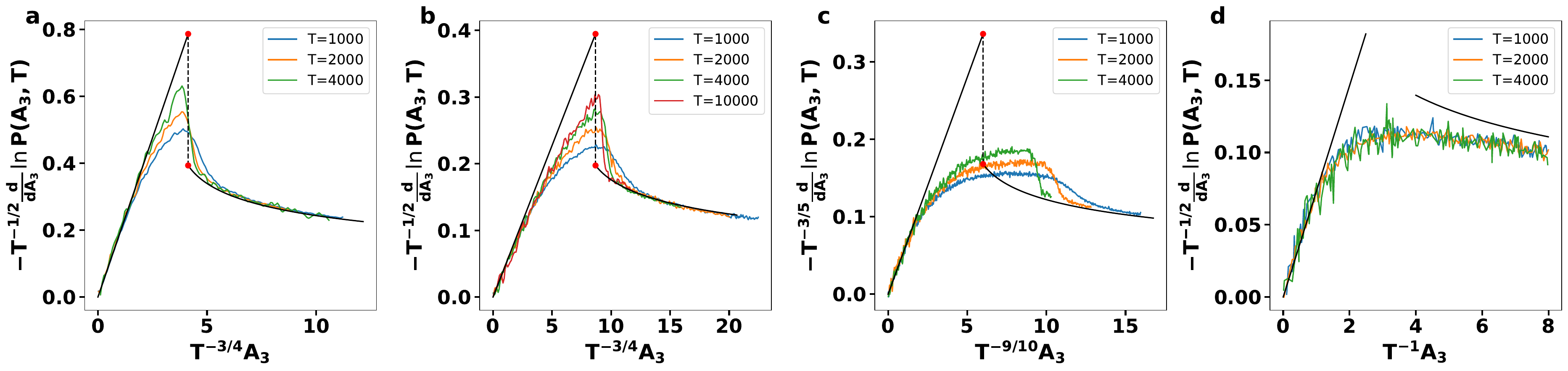}
\caption{The first derivative of the rescaled action, measured in the REWL simulations, with respect to $A_3$ for different Hurst exponents $H$: a) $H=1/4$; b) $H = 1/2$; c) $H=3/5$, and d) $H=3/4$. The red dots in the panels a-c mark the finite discontinuity of the derivative. }
\label{fig_RFD}
\end{figure}

One can see from Figs.~\ref{fig_2} and \ref{fig_RF} that, as $T$ increases, the simulation results slowly converge toward the analytical predictions, and the DPT in the panels a-c  becomes more pronounced. The convergence is better seen in the plots of the numerical derivative of the rescaled simulated actions with respect to $A_3$, shown in the right panel of Fig.~\ref{fig_2} and in Fig.~\ref{fig_RFD}. The derivative plots highlight a gradual sharpening of the transition with an increase of $T$ and confirms the anomalous scaling exponents characterizing the emerging DPT.

\subsubsection{Optimal paths}

\begin{figure}[ht]
\centering
\includegraphics[clip,width=0.9\textwidth]{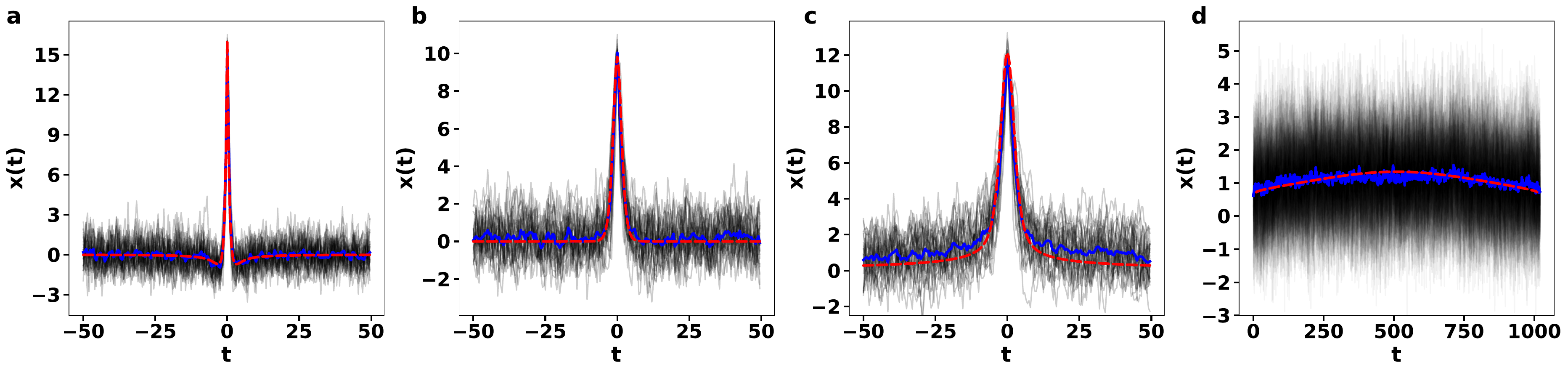}
\caption{$N$ realizations of the fOU process $x(t)$  conditioned on $A_3$  for $T=10^3$ (the black lines). The blue lines correspond to the trajectories averaged over the $N$ realizations. The red dashed lines show theoretically predicted optimal paths.  Panel a: $H=1/4$, $N = 30$, $A_3 =y T^{3/4}$ for $y=15$. Panel b: $H=1/2$, $N = 30$, $A_3 =y T^{3/4}$ for $y=15$. Panel c:  $H=3/5$, $N = 30$, $A_3 =y T^{9/10}$ for $y=5$. Panel d: $H=3/4$, $N = 100$, $A_3 =y T$ for $y=8$.}
\label{fig_6}
\end{figure}

We also sampled, in the REWL simulations, individual realizations of the fOU process $x(t)$ conditioned on different values of $A_3$ in the mixed region above the predicted phase transition. Examples of such realizations, clearly showing instanton configurations, are presented in Fig. \ref{fig_6}. The black curves represent the sampled trajectories, arbitrarily shifted in time such that the instanton maxima are positioned at $t=0$. The blue lines show the ensemble-averaged trajectories. The red dashed lines represent the optimal paths obtained by the numerical solution of the nonlinear integral equation~(\ref{IntEqkappa}), as explained in Sec. \ref{Sec4a}.  In the mixed scenario with a DPT, that is for $H=1/4$, $H=1/2$, and $H=3/5$, the optimal paths correspond to the instanton contribution, given by the minimizer $z_*(y)$ of Eq.~(\ref{F}). As one can see from Fig. \ref{fig_6}, the instanton, which gives the predicted contribution to the rate function, agrees well with the  ensemble-averaged trajectory. For $H=3/4$ the optimal path is delocalized, and the measured trajectories are plotted without any shifts. In this case the optimal path corresponds to the value of $A^\text{deloc}_3$ matching that of the ensemble-averaged trajectory $A^\text{deloc}_3 = \int\limits_0^T \langle x\rangle^3_\text{ens}(t)dt$.

\begin{figure}[ht]
\centering
\includegraphics[clip,width=0.9\textwidth]{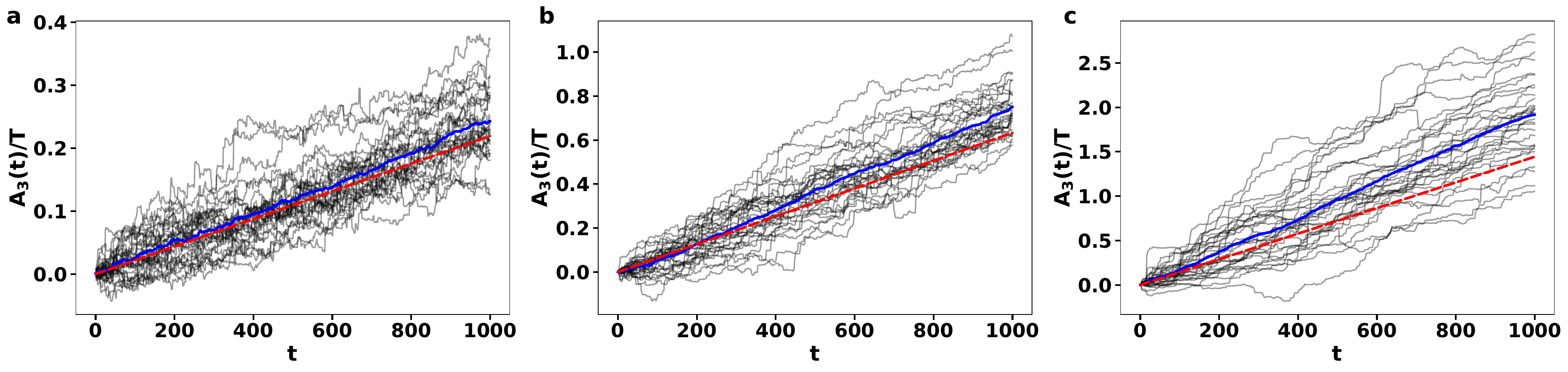}
\caption{$N =30$ realizations of $A_3(t)$. The blue lines correspond to the trajectories averaged over the $N$ realizations. The red dashed lines show the theoretically predicted uniform contributions for: a) $H=1/4$, $A_3 =y T^{3/4}$ for $y=15$; b) $H=1/2$, $A_3 =y T^{3/4}$ for $y=15$; and c):  $H=3/5$,  $A_3 =y T^{9/10}$ for $y=15$.}
\label{fig_7}
\end{figure}

In addition to the instanton contribution to $A_3$, we also examined the non-instanton part, coming from the fluctuations uniformly distributed along the trajectory. Specifically, we computed the dynamical observable itself at intermediate times $0\leq t\leq T$:
\begin{equation}
    A_3(t)=\int\limits_{0}^t x_\text{uniform}(s)^3ds,
\end{equation}
where each trajectory $x_\text{uniform}(s)$ is obtained by subtracting the corresponding measured optimal path  (shown in Fig.~\ref{fig_6}) from the sampled trajectory $x(s)$, and the optimal path is aligned with the instanton maxima location of the trajectory. The black curves in Fig.~\ref{fig_7} represent  the dynamical observables $A_3(t)$ at intermediate times, calculated using the same ensemble of trajectories as in Fig.~\ref{fig_6}. The blue lines show the ensemble-averaged values of $A_3(t)$. The red lines represent the theoretically predicted uniform contribution $A_3-A_{3,\text{inst}}$. As one can see, a better agreement between the theoretical and simulated slopes, and between the theoretical optimal paths and the sampled instanton configurations in Fig.~\ref{fig_6}, is observed for smaller $H$. This observation, which is to be expected, correlates with the fact that a faster convergence of the simulated action to the theoretical one in Figs.~\ref{fig_2} and \ref{fig_RF} is also observed for smaller $H$.

\section{Summary and Discussion}
\label{discussion}
As can be seen  from the phase diagrams, shown in Figs. \ref{fig_PD_alpha_n} and \ref{fig_PD_H_n}, the long-time large-deviation statistics of $A_n$, $n>2$, of stationary Gaussian processes are strongly affected by the correlation properties of the process or, more precisely, by the long-time behavior of the autocovariance $\kappa(t)$. 
Not very surprisingly, all short-correlated Gaussian processes behave in this respect very similarly to the previously studied 
Ornstein-Uhlenbeck process \cite{Smith2022}. That is, their probability densities $P(A_n,T)$ share the same anomalous scaling behavior and the same type of the first-order DPT in the limit of $T\to \infty$.  

Gaussian processes with longer correlations behave differently, and the differences grow as the correlations become more long-ranged. For sufficiently heavy correlation tails the large-deviation function of $A_n$ becomes smooth: no DPT is observed in the limit of $T\to \infty$. The reason for the analytical behavior of the rate function in this regime is the non-existence of strongly localized solutions for the optimal paths of the process conditioned on large $A_n$.   

Remarkably, there is also an intermediate region of moderately-correlated processes. Here a first-order DPT is present at $T\to \infty$. The long-time scaling behavior of  $P(A_n,T)$, however is different in this case from that for the short-correlated processes, as it explicitly depends on the exponent of the power-law decay of $\kappa(t)$ at $t\to \infty$.  The key factor behind the new scaling regime is the effect of long correlations on the typical, Gaussian fluctuations of $A_n$. 

The localized-in-time instanton-like optimal path, which plays a crucial role in the anomalous scaling and the condensation transition, is qualitatively similar to the ``big jump"  
which dominates the far tail of the distribution $P(S_N,N)$ of the sum $S_N = \sum_{i=1}^N y_i$ of independent identically distributed random variables $y_i$ in the large-$N$ limit. The big-jump principle (BJP) states that, for subexponential tails of the  distribution of the random variables, $-\ln P(y_i\to \infty)\sim y_i^{\epsilon}$ with $0<\epsilon<1$, the distribution tail of the $S_N$ is asymptotically dominated by the single largest summand \cite{Nagaev,Petrov,Denisov,Barkai2}. Discretizing the time integral in Eq.~(\ref{An}), one finds that, at $n>2$,  the (correlated!) random variables  $y_i\equiv x^n_i$ in our problem do have a subexponential tail with the power $\epsilon=2/n<1$.  Therefore, one can expect that, for short-correlated processes, the BJP must hold. Essentially, we have shown here that this is indeed what happens, and have also examined what happens when the correlations increase until the BJP breaks down.

\bigskip
\noindent
{\bf Acknowledgment}. We are very grateful to Naftali R. Smith for his insightful advice and suggestions. This research was supported by the Israel Science Foundation (Grant No. 1579/25). A.V. is supported by Instituto Balseiro and by a Simons Foundation Targeted Grant to Institutions.

\end{document}